\documentclass{elsart1p}
\usepackage{graphicx}
\usepackage{amssymb}
\begin{document}
\begin{frontmatter}
\title{Nuclear symmetry energy: An experimental overview}
\author{D.V. Shetty}
\address{Physics Department, Western Michigan University, Kalamazoo, MI 49008-5252, USA}
\author{S.J. Yennello}
\address{Cyclotron Institute, Texas A$\&$M University, College Station, TX 77843, USA \\*Chemistry Department, Texas A$\&$M University, College Station, TX 77843, USA}

\begin{abstract}
The nuclear symmetry energy is a fundamental quantity important for studying the structure of systems as diverse as the atomic nucleus and the neutron star. Considerable efforts are being made to experimentally extract the symmetry energy and its dependence on nuclear density and temperature. In this article, we review experimental studies carried out up-to-date and their current status.
\end{abstract}

\end{frontmatter}

\section{Introduction}
The nuclear matter symmetry energy, which is defined as the difference in energy per nucleon between the pure neutron matter and the symmetric nuclear matter, is an important quantity that determines the properties of  objects such as  the atomic nucleus and the neutron star \cite{LI08}. The study  of symmetry energy and its  dependence on nuclear density and temperature is currently a subject of great interest \cite{BAR05}. Theoretically, the symmetry energy can be determined from microscopic calculations such as the Brueckner-Hartree-Fock (BHF) and the Dirac-Brueckner-Hartree-Fock (DBHF) calculations, or the phenomenological calculations such as the Skyrme Hartree-Fock (SHF) and the relativistic mean field (RMF) calculations \cite{LI08}. These calculations currently predict wide range of symmetry energies for densities below and above normal nuclear density, $\rho_{o}$ = 0.16 fm$^{-3}$ (Fig. 1). Experimentally, the symmetry energy is not a directly measurable quantity and has to be extracted indirectly from observables that are related to the symmetry energy. The experimental determination of the symmetry energy is therefore dependent on how reliable the model that describes the experimental observable is. So far there has been very few experimental determination of the symmetry energy. These experimental studies are of two types:
\begin{enumerate}
\item{} where, a certain form of the density dependence of the symmetry energy is assumed in the theoretical calculation and the experimental observable reproduced using the dependence that best explains the data. Such studies \cite{SHE04,CHE05,LI05,SHET07,SHE07,FAM06,GAL09,TSA09} often make use of dynamical models such as, the Isospin Boltzmann Uehling Uhlenbech (IBUU04) \cite{LI04}, the Improved Quantum Molecular Dynamics (ImQMD) \cite{ZHA08}, the Boltzmann Nordheim Vlasov (BNV) transport calculation \cite{BAR05} and the Anti symmetrized Molecular Dynamics (AMD) \cite{ONO03}, to relate the symmetry energy to the experimental observable. The disadvantage in such studies is that they assume a single form of the symmetry energy for all densities, which may not be true. Such studies  provide only a gross dependence of the symmetry energy  without much insight into its evolution with the density. Also, these studies are highly model dependent as will be discussed in a later section.
\item{} where,  the form of the density dependence of the symmetry energy is not known a priori and the symmetry energy is studied by mapping its value at each density. Such studies \cite{SHE07,SHE09,KOW07} require a detailed understanding of the relation between the symmetry energy, excitation energy, density and temperature.  They make use of the Statistical Multifragmentation Model (SMM) \cite{BOT02,TSA01}  to relate the symmetry energy to the experimental observable. The advantage in such studies is that they  provide a ``direct"  means of studying the symmetry energy with density and temperature that can then be compared with any theoretical predictions. The difficulty in such studies is often the theoretical interpretation of the symmetry energy.
\end{enumerate}

\begin{figure}
\begin{center}
\includegraphics[width=5.0 in, height=2.2 in]{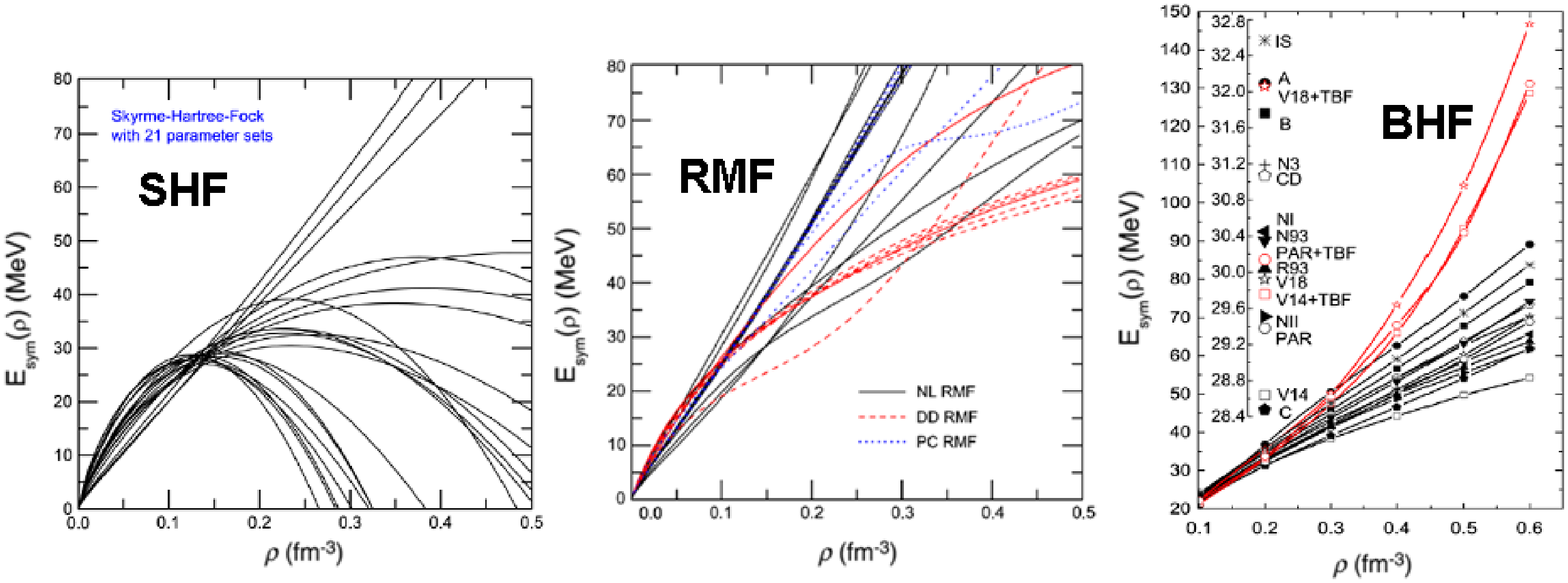}
\caption{Density dependence of the symmetry energy as predicted by various theoretical calculations [1].}
\end{center}
\end{figure}

In this article, we review recent experimental studies on the symmetry energy and its density dependence obtained using the above two approaches.

\section{Nuclear matter symmetry energy below saturation density ($0.3$ $\rho_o$ $\le$ $\rho$ $\le$ $\rho_o$)}
The symmetry energy below saturation density has been studied  in  heavy ion multifragmentation reactions using fragments with A $>$ 4. Observables such as the fragment yield, isoscaling parameter, isospin diffusion transport ratio, double neutron-proton ratio, pre-equilibrium emission, and $\langle$N$\rangle$/Z has been used to extract the symmetry energy.

\subsection{Symmetry energy from dynamical multifragmentation model comparison}
The symmetry energy in this density region has been studied by comparing the experimental observables with the dynamical model calculations. Chen {\it {et al.}} \cite{CHE05}, compared the NSCL-MSU isospin diffusion data from $^{124}$Sn + $^{112}$Sn reaction at 50 MeV/A with the dynamical IBUU04 \cite{LI04} calculation, and obtained a symmetry energy of the form, E$_{sym}(\rho)$ = 31.6($\rho$/$\rho_{o}$)$^{\gamma}$ with $\gamma$ = 1.05.  The NSCL-MSU isospin data was also compared with the IBUU04 by Li {\it {et al.}} \cite{LI05},  that included the isospin dependence of the in-medium nucleon-nucleon cross-section, giving symmetry energy of the form, E$_{sym}(\rho)$ = 31.6($\rho$/$\rho_{o}$)$^{\gamma}$ with $\gamma$ = 0.69. Shetty {\it {et al.}} \cite{SHE04,SHET07,SHE07},   extracted the symmetry energy by comparing the isoscaling parameters from $^{40}$Ar, $^{40}$Ca + $^{58}$Fe, $^{58}$Ni and $^{58}$Fe, $^{58}$Ni +  $^{58}$Fe, $^{58}$Ni reactions with the dynamical AMD model calculation \cite{ONO03}, resulting in symmetry energy, E$_{sym}(\rho)$ = 31.6($\rho$/$\rho_{o}$)$^{\gamma}$ with $\gamma$ = 0.69. Famiano {\it {et al.}} \cite{FAM06}, studied the symmetry energy by comparing the experimental double neutron to proton ratio  in $^{112}$Sn + $^{112}$Sn and $^{124}$Sn + $^{124}$Sn reactions with the BUU97 calculation \cite{LI97}  and  obtained a symmetry energy, E$_{sym}(\rho)$ = 32($\rho$/$\rho_{o}$)$^{\gamma}$ with $\gamma$ = 0.5. Galichet {\it {et al.}} \cite {GAL09}, studied the symmetry energy by comparing the isospin diffusion and pre-equilibrium emission data in $^{58}$Ni + $^{58}$Ni, $^{197}$Au reactions at 52 - 74 MeV/A with the BNV calculation, and obtained a symmetry energy that increase linearly with density.  More recently, Tsang {\it {et al.}} \cite{TSA09}, compared the isospin diffusion and the neutron to proton double ratio for $^{124}$Sn + $^{112}$Sn reaction at 50 MeV/A with the ImQMD calculation \cite{ZHA08} and obtained a symmetry energy of the form,  E$_{sym}(\rho)$ = 12.5($\rho$/$\rho_{o}$)$^{2/3}$ + 17.6($\rho$/$\rho_{o}$)$^{\gamma}$ with $\gamma$ = 0.4 - 1.05. The functional dependence of the symmetry energy obtained from all these comparisons are very similar to each other.

It must be mentioned that while both the IBUU04 and ImQMD comparisons lead to similar form of the density dependence of the symmetry energy for the isospin diffusion observable, the IBUU04 fails to adaquately  explain  the neutron to proton ratio \cite{LI06}. Furthermore, these comparisons lead to very different results for densities above saturation density \cite{XIA09,FEN10} (as discussed in section 5). The determination of the symmetry energy from the dynamical model comparison is therefore highly model dependent, and the assumption of a single form of the symmetry energy at different densities questionable. 

\subsection{Symmetry energy from statistical multifragmentation model comparison}
The symmetry energy has also been studied  by comparing the experimental observables with the statistical multifragmentation model (SMM) calculation \cite{BOT02} in this density region. In multifragmentation a nucleus expands with increasing excitation energy and its equilibrium density is reduced from the ground state density. Such a decrease should essentially map the density dependence of the symmetry energy.  A detailed understanding of the relationship between excitation energy, density, and temperature is however required. An attempt has recently been made by Shetty {\it {et al.}} \cite{SHE07,SHE09} from the study of isoscaling parameters in $^{58}$Fe, $^{58}$Ni +  $^{58}$Fe, $^{58}$Ni reactions at 30, 40 and 47 MeV/A. In this work, the symmetry energy was studied by correlating  the experimentally observed decrease in the isoscaling parameter, density, and the flattening of temperature (caloric curve), with the increase in excitation energy. 

The decrease in the symmetry energy with increasing excitation energy was also observed in the fragmentation of excited target residues following $^{12}$C + $^{124}$Sn, $^{112}$Sn reactions at 300 and 600 MeV/A by Le Fevre {\it {et al.}} \cite{FEV05}.  It was also observed in $^{40}$Ar, $^{40}$Ca + $^{58}$Fe, $^{58}$Ni reactions at 25, 33, 45 and 53 MeV/nucleon by Iglio {\it {et al.}} \cite{IGL06}, and by Shetty {\it {et al.}} \cite{SHE04,SHE05}. Although in these work, a detailed understanding of the relation between temperature, excitation energy and nuclear density was not undertaken. Recently, Ogul {\it {et al.}} \cite{OGU09}, compared the MSU experimental isoscaling data ($\alpha$ = 0.36 and $\beta$ = -0.39) from $^{112}$Sn, $^{124}$Sn + $^{112}$Sn reaction with the SMM calculation, and observed that a significant reduction of the symmetry energy is necessary to reproduce the experimental data. Geraci {\it {et al.}} \cite{GER07}, studied $^{124}$Sn + $^{64}$Ni and $^{112}$Sn + $^{58}$Ni  reactions at 35 MeV/A and observed a similar decrease in the symmetry energy with increasing excitation energy. 

The reduction in the symmetry energy has also been observed in the projectile fragmentation of $^{64}$Ni + $^{64}$Ni, $^{86}$Kr + $^{64}$Ni, $^{124}$Sn, $^{208}$Pb reactions at 25 MeV/A using the observable $\langle$Z/A$\rangle$ by Souliotis {\it {et al.}} \cite{SOU06,SOU07}.  More recently, Hudan {\it {et al.}} \cite{HUD09}, studied the projectile fragmentation in $^{124}$Xe + $^{112}$Sn reaction at E/A = 30 MeV and measured the $\langle$N$\rangle$/Z and isotope distribution of fragments with Z $>$ 6. They observed that a reduced value of the symmetry energy from 25 MeV to 14 MeV is essential for explaining the data. They further showed that fragments with Z $>$ 6 are even more sensitive to the variation in symmetry energy and can be an important probe for studying symmetry energy.  A decrease in the symmetry energy was also implied by  Wuenschel {\it {et al.}} \cite{WUE09}, in the study of $^{86,78}$Kr + $^{64,58}$Ni reactions at 35 MeV/A. 

It has been argued \cite{SOUZ08} that the decrease in the symmetry energy to reproduce the experimental observables  in these studies may not be necessary if the surface corrections to the symmetry energy are included in the mass parameterization used in the SMM model calculation. The possibility of modification of the symmetry energy and the surface energy coefficient of nuclear matter was studied recently by Ogul {\it {et al.}} \cite{OGU09}. They found that the isoscaling parameters are affected very little by the surface energy variation and are very sensitive to the symmetry energy.  A decrease in the symmetry energy is therefore essential  to explain the experimental data.  W. Ye {\it {et al.}} \cite{YE09}, also studied the influence of the surface entropy on the isoscaling using the Extended Compound Nucleus (ECN) model and found that although the surface entropy increases the numerical values of isoscaling parameters it has only minor effect on the extracted symmetry energy. A clear theoretical interpretation of the symmetry energy in statistical model is therefore important. 

\subsection{Symmetry energy from other studies}
The nuclear symmetry energy has also been studied from observables other than those studied using heavy ion reactions. Centelles {\it {et al.}} \cite{CEN09}, have studied the symmetry energy using the experimental neutron skin measured in 26 antiprotonic atoms. Klimkiewicz {\it {et al.}} \cite{KLI07}, have studied the symmetry energy from correlation between the symmetry pressure and the symmetry energy using the properties of Pygmy Dipole Resonance (PDR) in $^{208}$Pb nuclei. Trippa {\it {et al.}} \cite{TRI08}, have studied the symmetry energy from the correlation between the symmetry energy and the experimental centroid energy of the Giant Dipole Resonance (GDR) in $^{208}$Pb nuclei. Khoa {\it {et al.}} \cite{KHO05,KHO07}, have studied the symmetry energy from the folding model analysis of the charge exchange $p$($^{6}$He, $^{6}$Li)$n$ reaction measured at 41.6 MeV/A. Danielewicz \cite{DAN03} have studied the symmetry energy by constraining the binding energy, neutron skin and the isospin analogue state of finite nuclei. The symmetry energy has also been studied by fitting the binding energy of 1654 nuclei using the Thomas-Fermi model of Myers and Swiatecki \cite{MYE96,MYE98}. The results of these independent studies are shown in Fig. 2 and 3 and discussed in section 4.

\section{Nuclear matter symmetry energy below $\rho$ $<$ $0.3$ $\rho_o$}
Symmetry energy at very low densities (0.01 - 0.05) $\rho_o$  has been studied by Kowalski {\it {et al.}}\cite{KOW07}, in heavy ion collision of $^{64}$Zn  + $^{92}$Mo, $^{197}$Au reactions at 35 MeV/A. It was shown that isoscaling analysis of the light clusters (A $\le$ 4) can be used to study the symmetry energy at such low densities. They observed that the experimental symmetry energy is somewhat higher than those expected from the mean field calculations. At such low densities cluster formation becomes important and mean field calculations do not take into account such effect.  Recently, Natowitz {\it {et al.}} \cite{NAT10}, have shown that the symmetry energy at such low densities can be explained by quantum statistical calculation that includes cluster correlation in  nuclear medium.  At densities higher than 0.2 - 0.3 $\rho_o$ the many body correlation disappears and the symmetry energy follows the dependence predicted by the mean field calculations.    

\section{Current status of the density dependence of the symmetry energy for ($\rho$ $\le$ $\rho_o$) }
The symmetry energy obtained from the studies discussed above are as shown in Fig. 2. The orange box on the extreme left  of fig. 2 correspond to the values of symmetry energy extracted by Kowalski {\it {et al.}} \cite{KOW07,NAT10}. The green and the red solid points are the symmetry energies extracted from the  correlation between  temperature, density, isoscaling parameter and the excitation energy by Shetty {\it {et al.}} \cite{SHE09,SHE07}. The blue solid point in the figure correspond to the symmetry energy obtained by constraining the experimental energy of giant dipole resonance (GDR) in $^{208}$Pb by Trippa {\it {et al.}} \cite{TRI08}. The square solid point in the figure correspond to the symmetry energy obtained by fitting the experimental differential cross-section data in a charge exchange reaction using the isospin dependent interaction of the optical potential by Khoa {\it {et al.}} \cite{KHO05,KHO07}.  The green solid curve in the figure is the one obtained from the dynamical AMD model comparison of the experimental isoscaling data  assuming the sequential decay effect to be small \cite{SHE04,SHET07}. The symmetry energy extracted from the IBUU04 comparison of isosopin diffusion results by Li {\it {et al.}}  \cite{LI05}, and the ImQMD comparison by Tsang {\it {et al.}} \cite{TSA09}, are similar to the green curve.  The solid black curve is the symmetry energy extracted from the double neutron to proton ratio by Famiano {\it {et al.}} \cite{FAM06}, using the BUU97 model calculation. The red dashed curve corresponds to the one obtained from an accurately calibrated relativistic mean field calculation by Todd$-$Rutel and Piekarewicz \cite{TOD05} for describing the giant monopole resonance (GMR) in $^{90}$Zr and $^{208}$Pb, and the isovector giant dipole resonance (IVGDR) in $^{208}$Pb. The shaded region in the figure corresponds to those obtained by constraining the binding energy, neutron skin thickness and isospin analogue state in finite nuclei using the mass formula by Danielewicz \cite{DAN03}. The yellow curve correspond to the parametrization adopted in the studies of neutron star \cite{HEI00}. Current studies  in the region, 0.3 $\rho_o$ $\le$ $\rho$ $\le$ $\rho_o$, can therefore be parameterized by a ``stiff" form of the symmetry energy, E$_{sym}(\rho)$ = 31.6$(\rho/\rho_{o})^{\gamma}$, with $\gamma$ = 0.5 - 0.7. This parameterization is in good agreement with the mean field calculation. In the density region, $\rho$ $<$ $0.3$ $\rho_o$, the experimental symmetry energy is observed to deviate from the mean field calculation due to the importance of cluster formation, and is in good agreement with the quantum statistic calculation \cite{NAT10}.

The slope parameter, L = 3$\rho_o$ ($\partial E_{sym}(\rho)/\partial \rho$),  is an alternate way of constraining the symmetry energy near saturation density, $\rho_o$.  Fig. 3 shows the slope parameter L, obtained from various studies discussed above. These comparison results in  a constraint of  30 $<$ L $<$ 80 MeV for the symmetry energy slope at saturation density \cite{CEN09,CHO09}. 

\section{Nuclear matter symmetry energy above saturation density ($\rho$ $>$ $\rho_o$) }
At densities higher than the normal nuclear matter density the theoretically determined symmetry energy as a function of density is largely unconstrained. Experimental determination of the symmetry energy at such densities  have been very few and  they are all from dynamical model comparison. The results are highly model dependent and contradicting.  
\begin{figure}
\begin{center}
\includegraphics[width=3.2 in,height=3.2 in]{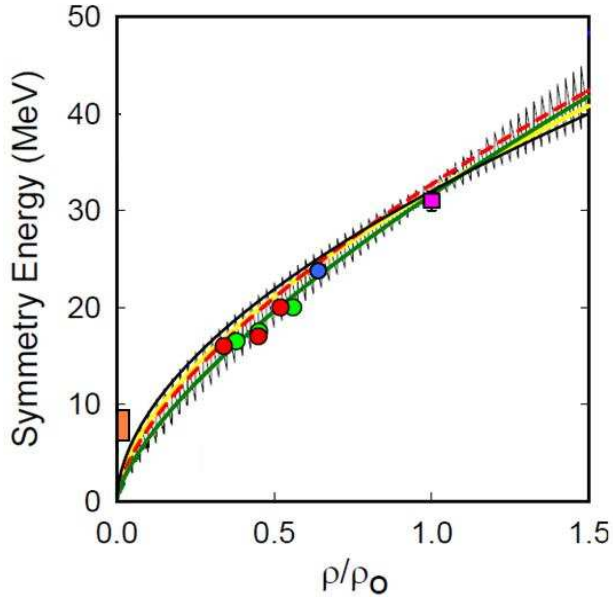}
\caption{Density dependence of the symmetry energy extracted from various different studies as described in the text.}
\end{center}
\end{figure}
\begin{figure}
\begin{center}
\includegraphics[width=3.5 in, height=3.5 in]{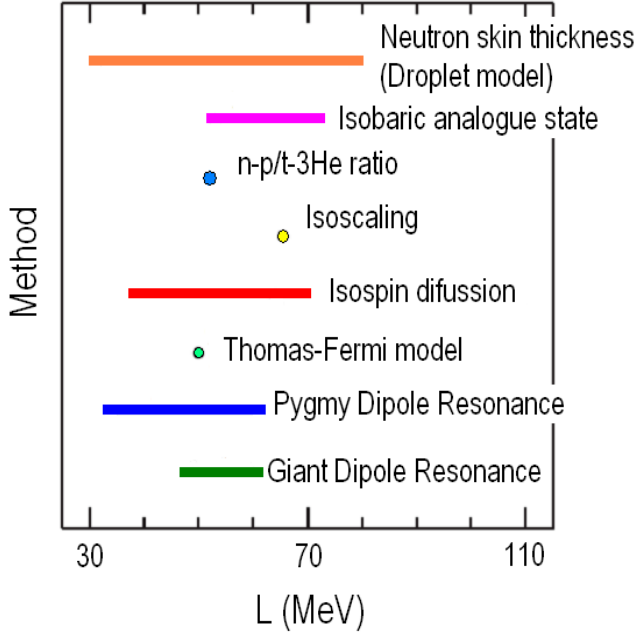}
\caption{The slope parameter L, from various different observables as described in the text.}
\end{center}
\end{figure}
 One  observable that has recently been studied  is the $\pi^{-}$$/$$\pi^{+}$ ratio in $^{40}$Ca + $^{40}$Ca, $^{96}$Ru + $^{96}$Ru, $^{96}$Zr + $^{96}$Zr and $^{197}$Au + $^{197}$Au reactions at GSI by the FOPI collaboration \cite{REI07}. A comparison \cite{XIA09} of this data with the transport model IBUU04 calculation shows that it  can be reproduced with a ``soft" (x = 1)  form of the density dependence of the symmetry energy. This is in contrast to those obtained from the low density studies of the symmetry energy where a ``stiff" (x = 0) form of the symmetry energy reproduced the data well.  More recently, the $\pi^{-}$$/$$\pi^{+}$ GSI data was also compared with the ImQMD calculation by Feng {\it {et al.}} \cite{FEN10}. The comparison favored a much stiffer form of the symmetry energy than those obtained from the low density studies, and in direct contradiction to those obtained from the IBUU04 model comparison. The symmetry energy results at such high densities is therefore currently controversial.

Another observable that has been suggested for probing the high density behavior of the symmetry energy is the relative and differential collective flow between triton and $^{3}$He particles \cite{YON09}. Preliminary calculation by Yong {\it {et al.}} \cite{YON09}, suggests that a ``stiff" (x = 0) dependence of the symmetry energy would lead to a linear behavior for the relative flow, whereas the``soft"  (x = 1) dependence would give a non linear behavior with appreciable difference between the two. So it is expected that a measurement of this observable should provide important information on the high density behavior of the symmetry energy.

The theoretical determination of the symmetry energy at high density is a challenge. It is not known how reliable the dynamical model calculations can be at such high densities as  it is not known how the quantum many body effects can be treated in these calculations, or how the effects of spin$-$isospin for three body force can be included. Also, the nuclear matter at such densities is not in a pure nucleonic state  at thermodynamical equilibrium.  The above comparison with the experimental data is therefore only a ``circumstantial'' evidence \cite{XIA09} and much theoretical work needs to be done before a definitive conclusion can be reached.  

\section{How is the nuclear matter symmetry energy related to the symmetry energy of finite nuclei ?}
The symmetry energy of finite nuclei at saturation density is often extracted by fitting ground state masses with various versions of the liquid drop mass formula. For a finite nuclei it is important to decompose the symmetry term of liquid into bulk (volume) and surface contributions, and identify the volume symmetry energy coefficient as the symmetry energy derived from the nuclear matter at saturation density. Using the constraint obtained from the above studies on  nuclear matter symmetry energy, the symmetry energy of a finite nucleus of mass $A$,  $S_{A}(\rho)$, can be written as \cite{SHE09},

\begin{equation}
  S_{A}(\rho) = \frac{\alpha(\rho/\rho_{\circ})^{\gamma}}{1 + [\alpha(\rho/\rho_{\circ})^{\gamma}/\beta A^{1/3}]}
\end{equation}

where, $\alpha$ = 31 - 33 MeV, $\gamma$ = 0.5 - 0.7 and $\alpha/\beta$ = 2.6 - 3.0. The quantities $\alpha$ and $\beta$ are the volume and the surface symmetry energy at normal nuclear density. Presently, the values of  $\alpha$, $\gamma$ and $\alpha/\beta$ remain unconstrained. The ratio of the volume symmetry energy to the surface symmetry energy ($\alpha/\beta$), is closely related to the neutron skin thickness \cite{DAN03}. Depending upon how the nuclear surface and the Coulomb contribution is treated, two different correlations between the volume and the surface symmetry energy have been predicted \cite{STE05} from fits to nuclear masses. Experimental masses and neutron skin thickness measurements for nuclei with $N/Z$ $>$ 1 should provide tighter constraint on the above parameters.

The above relation for the  symmetry energy of finite nuclei has been compared \cite{SHE09} with the Thomas-Fermi calculation of Samaddar {\it {et al.}} \cite{SAM08}. It was observed that this empirical relation compares very well with the Thomas-Fermi calculation for wide range of nuclei. Future measurements of symmetry energy as a function of excitation energy for very light and heavy nuclei should provide further insight into the validity of this relation.

\section{Future measurements of symmetry energy at rare isotope beam facilities }
Currently, studies on the density dependence of the symmetry energy are being carried out using beams of stable nuclei. In order to study large isospin dependency beams of neutron-rich nuclei are needed. A number of observables sensitive to the symmetry energy  that can be experimentally tested has been sugggested by Di Toro {\it {et al}} \cite{BAR05,DIT08}. It is hoped that future facilities for rare isotope beams such as FRIB, FAIR and SPIRAL II should provide increased precision in the measurement of symmetry energy using these observables over a wide range of density. These facilities will allow studies to be carried out at 2 - 3 times the normal nuclear density. 

New detectors such as the Time Projection Chamber (TPC) at NSCL$/$MSU and the Time Projection Chamber, SAMURAI at RIKEN$/$Japan are therefore being planned to study the symmetry energy and the nuclear equation of state. These detectors will significantly enhance our understanding of the nuclear symmetry energy and their relation to atomic nuclei and neutron stars.

\section{Summary}
In summary, the experimental determination of the nuclear symmetry energy is reviewed. It is observed that despite the model dependent ways in which the symmetry energy is extracted  significant progress has been made. These studies are the first step  in an effort to constrain the symmetry energy that is  important for studying the structure of neutrons stars and exotic nuclei.  In the future, it is hoped that  more measurements will be carried out where the symmetry energy is  extracted for each densities and compared with theoretical calculations.  These  measurements  using beams of neutron rich nuclei and robust theoretical interpretation  of the symmetry energy will help refine some of these results. Furthermore, they will help constrain the symmetry energy at densities above normal nuclear density where experimental results are scarce.

\end{document}